%% file: conference_101719.tex
\documentclass[conference]{IEEEtran}
\IEEEoverridecommandlockouts
\usepackage{cite}
\usepackage{amsmath,amssymb,amsfonts}
\usepackage{algorithmic}
\usepackage{graphicx}
\usepackage{textcomp}
\usepackage{xcolor}

\usepackage{algorithmic}
\usepackage{graphicx}
\usepackage{textcomp}
\usepackage{xcolor}
\usepackage{multirow}
\usepackage{pifont}
\usepackage{hyperref}
\usepackage{tikz}
\usepackage{comment}
\usepackage{makecell}

\def\BibTeX{{\rm B\kern-.05em{\sc i\kern-.025em b}\kern-.08em
    T\kern-.1667em\lower.7ex\hbox{E}\kern-.125emX}}
\begin{document}

\title{Refactoring $\neq$ Bug-Inducing: Improving Defect Prediction with Code Change Tactics Analysis}

\author{
\IEEEauthorblockN{
Feifei Niu\IEEEauthorrefmark{1},
Junqian Shao\IEEEauthorrefmark{1},
Christoph Mayr-Dorn\IEEEauthorrefmark{2},
Liguo Huang\IEEEauthorrefmark{3},\\
Wesley K.G. Assunção\IEEEauthorrefmark{4},
Chuanyi Li\IEEEauthorrefmark{1},
Jidong Ge\IEEEauthorrefmark{1},
Alexander Egyed\IEEEauthorrefmark{2}
}

\IEEEauthorblockA{\IEEEauthorrefmark{1}Nanjing University, Nanjing, China\\}

\IEEEauthorblockA{\IEEEauthorrefmark{2}Johannes Kepler University, Linz, Austria\\}

\IEEEauthorblockA{\IEEEauthorrefmark{3}Southern Methodist University, Dallas, USA\\}

\IEEEauthorblockA{\IEEEauthorrefmark{4}North Carolina State University, Raleigh, USA\\}
}


\maketitle

\begin{abstract}
Just-in-time defect prediction (JIT-DP) aims to predict the likelihood of code changes resulting in software defects at an early stage. Although code change metrics and semantic features have enhanced prediction accuracy, prior research has largely ignored code refactoring during both the evaluation and methodology phases, despite its prevalence. Refactoring and its propagation often tangle with bug-fixing and bug-inducing changes within the same commit and statement. Neglecting refactoring can introduce bias into the learning and evaluation of JIT-DP models. 
To address this gap, we investigate the impact of refactoring and its propagation on six state-of-the-art JIT-DP approaches. We propose \underline{C}ode ch\underline{A}nge \underline{T}actics (CAT) analysis to categorize code refactoring and its propagation, which improves labeling accuracy in the JIT-Defects4J dataset by 13.7\%. Our experiments reveal that failing to consider refactoring information in the dataset can diminish the performance of models, particularly semantic-based models, by 18.6\% and 37.3\% in F1-score. Additionally, we propose integrating refactoring information to enhance six baseline approaches, resulting in overall improvements in recall and F1-score, with increases of up to 43.2\% and 32.5\%, respectively. Our research underscores the importance of incorporating refactoring information in the methodology and evaluation of JIT-DP. Furthermore, our CAT has broad applicability in analyzing refactoring and its propagation for software maintenance.
\end{abstract}

 \begin{IEEEkeywords}
Defect Prediction, Refactoring, Propagation
 \end{IEEEkeywords}

\section{Introduction} \label{sec:introduction}
Software defects are an inherent and inevitable aspect of the software development process~\cite{shull2000perspective}. If left unaddressed, these defects can lead to significant and immeasurable losses. For example, in 1994 in Scotland, a Chinook helicopter crashed because of a system error and killed all 29 passengers.\footnote{\url{https://encyclopedia.pub/entry/34354}}
Consequently, efficient defect handling remains a top priority among all software maintenance activities \cite{boehm2001defect}. In defect handling, software defect prediction (SDP) plays a crucial role in aiming to forecast the likelihood of defects at different granularities of source code, e.g., files, classes, and methods. Unlike bug localization, SDP's primary objective is to predict potential defects before any visible symptoms emerge, mitigating negative impacts on the software \cite{yan2020just}. 
This proactive approach empowers developers to implement remedial measures and enhance overall software quality. Furthermore, SDP enables the effective allocation of the testing effort and software quality assurance resources to focus on high-risk areas identified through its predictions~ \cite{pornprasit2022deeplinedp}. 
As a result, SDP has emerged as one of the most significant quality assurance activities \cite{kamei2012large}.

In recent years, significant advancements have been made in the field of SDP, with numerous studies proposing diverse techniques to address this critical problem~\cite{kamei2012large, rosen2015commit, shihab2012industrial, yang2016effort, yang2015deep, yang2017tlel, huang2017supervised, fu2017revisiting, yan2020just, ni2022best, pornprasit2021jitline, pornprasit2022deeplinedp}. Among these variants, Just-in-Time Defect Prediction (JIT-DP)~\cite{kamei2012large} stands out as a compelling approach that focuses on the early detection of software defects, particularly before the new version is released. JIT-DP operates by identifying code changes that are likely to introduce defects at the time of check-in. By classifying code changes as either \textit{buggy} or \textit{clean} during the submission process, JIT-DP offers several advantages. Foremost, JIT-DP supports timely correcting actions to be taken before significant losses can occur~\cite{kamei2012large, pornprasit2021jitline}. 
Additionally, JIT-DP leverages recent code changes and thus fosters accelerating defect resolution~\cite{yan2020just, kamei2012large} as developers retain the relevant development context in their minds. Moreover, from a research standpoint, accurate defect prediction is a crucial prerequisite for effective bug localization and repair. Only with precise defect prediction results can bug localization and repair efforts be carried out efficiently~\cite{yan2020just, ni2022best}. 

State-of-the-art (SOTA) JIT-DP techniques have primarily relied on code change metrics and semantic features, yielding promising results. \textit{Code change metrics} are defined by experts to characterize code changes based on their professional knowledge and experience. These metrics encompass diverse change-level factors, including diffusion, size, purpose, code change history, and developer expertise. 
Kamei et al.~\cite{kamei2012large} propose 14 code change metrics (in Table~\ref{tab:14metrics}), that have demonstrated effectiveness in predicting buggy commits \cite{pornprasit2021jitline, ni2022best, zeng2021deep, tabassum2020investigation}. \textit{Semantic features} are predominantly extracted from natural language text, such as change logs and commit messages, as well as source code, enabling the identification of semantic characteristics of the buggy and clean commits~\cite{hoang2019deepjit, hoang2020cc2vec, yang2015deep, aversano2007learning, kim2008classifying}.

Despite the plethora of solutions based on code change metrics and semantic features, we identified critical limitations in the SOTA JIT-DP approaches: \textbf{the oversight of code refactoring and its propagation, in both the evaluation and methodology phases}, which may cause bias to the evaluation results as well as restrict models' performance. \ding{202} During the evaluation process, the SOTA approaches are evaluated using datasets constructed with \textit{git blame} command or the SZZ algorithm family (original SZZ \cite{sliwerski2005changes}, and its variants: AG-SZZ \cite{kim2006automatic}, MA-SZZ \cite{da2016framework}, and RA-SZZ \cite{neto2018impact}) \cite{keshavarz2022apachejit, ni2022best, pornprasit2022deeplinedp}. To this end, \textbf{refactoring or its propagated code changes might be labeled as buggy in the evaluation datasets, while these changes should not induce bugs if implemented completely}. \ding{203} In SOTA approaches, neither the code change metrics nor the semantic features consider code refactoring and refactoring propagation. Since refactoring and its propagation are at low risk of inducing bugs (if implemented correctly), this oversight may limit the performance of these approaches. Exploring whether leveraging refactoring information can enhance model performance is a question worth investigating. 

In the literature, many refactoring detection tools have been proposed, e.g., RefDiff~\cite{silva2017refdiff, silva2020refdiff}, and RefactoringMiner~\cite{tsantalis2020refactoringminer, Tsantalis:ICSE:2018:RefactoringMiner}. The RA-SZZ algorithm~\cite{neto2018impact} employs the RefDiff~\cite{silva2017refdiff} approach to exclude refactored code lines from bug-inducing changes, thereby ensuring a more accurate evaluation dataset for JIT-DP. However, the RefDiff~\cite{silva2017refdiff} approach they employed identifies only 13 refactoring types, while there are over 90 types in existence. Most importantly, we have also identified two limitations in the existing refactoring mining tools that hinder their direct application to JIT-DP. Firstly, they only detect the presence of refactoring but do not untangle refactoring from other types of changes, while \textbf{a refactored statement may still induce bugs if it contains code changes beyond the refactoring itself}. Only purely refactored statements should be bug-free, which means it is necessary to untangle refactored code changes with other code changes in JIT-DP.
Secondly, \textbf{refactoring propagation is not detected by these tools}. Refactoring propagated changes are triggered by refactoring (for example, renaming a method will trigger changes in the statements that call this method), and if implemented correctly, they should not induce bugs. These two limitations can introduce significant bias into JIT-DP evaluation datasets.
Our statistics (in Section~\ref{sec:rq1}) indicate that a significant proportion of statements are mixed with refactoring and edit changes (false negative of refactored line, \textbf{15.8\%}), and there is also a portion of only propagated statements (false positive of refactored line, \textbf{8.3\%}) that have been overlooked by the refactoring detection tool \cite{neto2018impact, silva2017refdiff, silva2020refdiff, tsantalis2020refactoringminer, Tsantalis:ICSE:2018:RefactoringMiner}. To this end, the RA-SZZ algorithm still exhibits significant inaccuracies. Such inaccurate datasets can introduce bias into the results of experimental evaluations. However, to the best of our knowledge, there is currently no research exploring the extent of this bias on SOTA approaches. 

To understand how code refactoring and refactoring propagation influence SOTA approaches, in this paper, we firstly enhance existing refactoring detection tools by proposing our \underline{C}ode ch\underline{A}nge \underline{T}actics (CAT) analysis, especially refactoring propagation analysis, to discern purely refactored and its propagated code changes. Leveraging CAT, we augment the JIT-Defects4J \cite{ni2022best} dataset by correcting the label of 355 commits and carry out empirical evaluation on the SOTA approaches with the augmented dataset. Furthermore, we propose refactoring-aware metrics (RAMs) to enhance code change metrics-based approaches and apply refactoring information to augment semantic feature-based approaches.

\begin{figure}[htbp]
\centering
\includegraphics[width=0.9\linewidth]{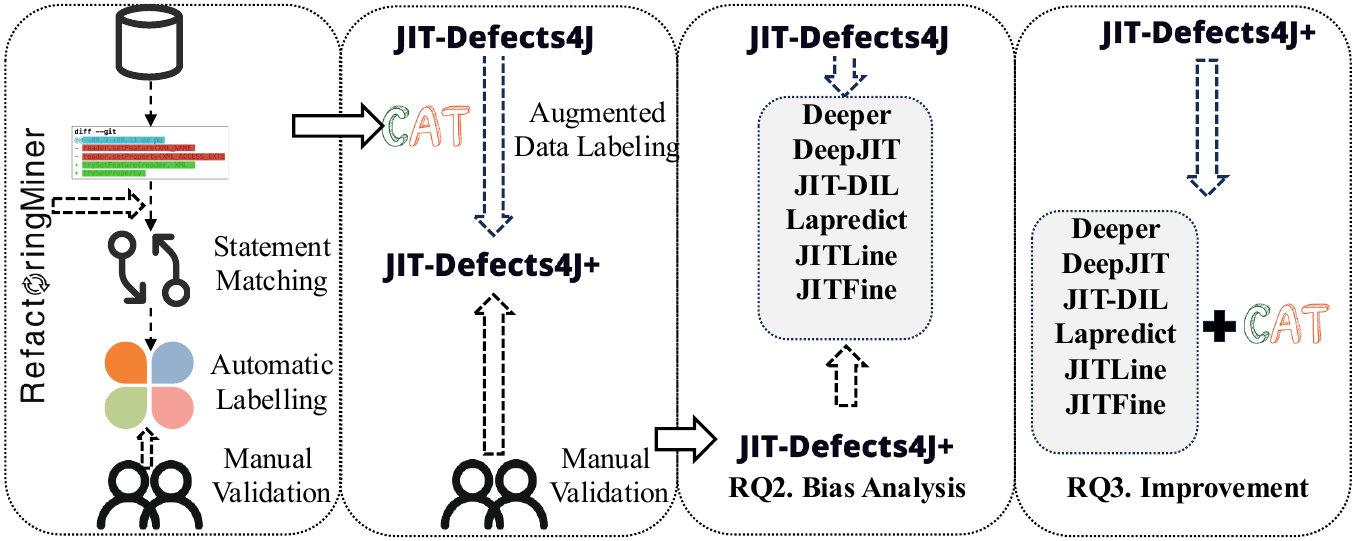}
\caption{Overall Framework.}
\label{fig:overall}
\end{figure}

The main contributions of this study are as follows (as shown in Fig.~\ref{fig:overall}):

\begin{itemize}
     
    \item We introduce and open-source CAT for labeling code change categories,\footnote{https://github.com/feifeiniu-se/CAT} which can be used to discern purely refactored lines (including moved lines) and lines affected by refactoring within code changes. 
    \item With CAT, we augment the JIT-Defects4J \cite{ni2022best} datasets by correcting the label of 355 commits. 
    
    \item Based on the augmented dataset, we empirically investigate the bias of evaluating SOTA JIT-DP approaches.
    \item We apply refactoring-related information labeled by CAT to both code change metrics-based approaches and semantic feature-based approaches, which can effectively improve the accuracy.
\end{itemize}

Our major empirical findings from this study include:
\begin{itemize}
    \item Code refactoring permeates throughout evolution history, with 41.3\% of commits containing at least one instance of refactoring, impacting 10.8\% of the modified lines. Additionally, there is a notable presence of purely refactoring propagation (8.3\%) and tangled refactoring (15.8\%) that cannot be disregarded in JIT-DP.
    
    \item In the bias analysis, four baseline methods (i.e., Deeper~\cite{yang2015deep}, DeepJIT~\cite{hoang2019deepjit}, JIT-DIL~\cite{yan2020just}, and LApredict~\cite{zeng2021deep}) demonstrated a high level of fault tolerance towards mislabeling in the dataset resulting from refactoring, with no significant differences observed. In contrast, the JITLine~\cite{pornprasit2021jitline} and JIT-Fine~\cite{ni2022best} approaches showed improved performance on the augmented dataset, suggesting that such datasets are more advantageous for models in acquiring accurate knowledge, thus enhancing their overall capabilities.

    \item Integrating the refactoring-related information generated by CAT into existing approaches led to improvements in all approaches (except for LApredict~\cite{zeng2021deep}), suggesting that refactoring-related information is highly effective and enhances the performance of each approach. This further underscores the overall effectiveness of our approach.
\end{itemize} 

This work is the first to separate refactoring and propagation code changes, the first to examine the influence of refactoring in JIT-DP, and the first to integrate refactoring into JIT-DP. Our research offers valuable insights for future investigations and serves as a foundational study.

The remainder of the paper is organized as follows. Section~\ref{sec:related} introduces the background and related work. Section~\ref{sec:motivation} motivates our work with two types of examples. Section~\ref{sec:approach} elaborates on CAT. Section~\ref{sec:experimental} describes the experimental setup. Section~\ref{sec:results} presents the empirical results. Section~\ref{sec:threats} discusses threats to validity. Section~\ref{sec:conclusion} concludes this paper.


\section{Background and Related Work} \label{sec:related}

\subsection{Code Change Metrics}\label{metrics}
\textit{Code change metrics} are firstly introduced by Kamei et al.~\cite{kamei2012large} based on prior proposed change measures, consisting of 14 types of code change factors with five dimensions, namely diffusion, size, purpose, code change history, and developer's experience, which are shown in Table~\ref{tab:14metrics}. Prior to that, Mockus and Weiss \cite{mockus2000predicting} showed that the number of touched subsystems is a good feature for detecting a defect. Hassan \cite{hassan2009predicting} also found that scattered changes are related to the probability of defects. Nagappan and Ball \cite{nagappan2006mining} and Moser et al. \cite{moser2008comparative} reveal that the size of code change (e.g., the number of lines of code added in a revision) is a good indicator of defect-prone modules. Yet, a bug-fixing commit is more likely to introduce a new defect \cite{guo2010characterizing, purushothaman2005toward}. These metrics have been adopted by many subsequent approaches and achieved promising results~\cite{ni2022best, pornprasit2021jitline, zeng2021deep, tabassum2020investigation}. Yang et al. use a Deep Belief Network (DBN) to learn features for defect prediction \cite{yang2015deep}. Yan et al. build a logistic regression-based model with 14 code change metrics \cite{yan2020just}. Zeng et al. leverage the number of added lines to learn defective features \cite{zeng2021deep}. While Pronprasit et al. \cite{pornprasit2021jitline} as well as Ni et al. \cite{ni2022best} both combined 14 code change metrics with semantic features for predicting defects.
 
\begin{table}[]
\caption{Original 14 basic change-level features}
\label{tab:14metrics}
\addtolength{\tabcolsep}{-5pt}
\begin{tabular}{|l|l|c|}
\hline
\textbf{Name}    & \textbf{Description}  & \textbf{Dimension} \\ \hline \hline
NS      & The number of modifed subsystems \cite{mockus2000predicting}    & \multirow{4}{*}{Diffusion}     \\
ND      & The number of modifed directories \cite{mockus2000predicting}   &    \\
NF      & The number of modifed files \cite{nagappan2006mining}        &    \\
Entropy & Distribution of modifed code across each file \cite{d2010extensive, hassan2009predicting}        &    \\ \hline \hline
LA      & Lines of code added \cite{moser2008comparative, nagappan2005use}  & \multirow{3}{*}{Size} \\
LD      & Lines of code deleted \cite{moser2008comparative, nagappan2005use}     &    \\
LT      & Lines of code in a file before the change \cite{koru2008investigation}      &    \\ \hline \hline
FIX     & Whether or not the change is a defect fix \cite{guo2010characterizing, purushothaman2005toward}      & \multicolumn{1}{c|}{Purpose}   \\ \hline \hline
NDEV    & \begin{tabular}[c]{@{}l@{}}The number of developers that changed \\ the modifed files \cite{matsumoto2010analysis}\end{tabular}       & \multirow{3}{*}{History}       \\
AGE     & \begin{tabular}[c]{@{}l@{}}The average time interval between the last \\ and current change \cite{graves2000predicting}\end{tabular} &    \\
NUC     & \begin{tabular}[c]{@{}l@{}}The number of unique changes to the \\ modifed files \cite{d2010extensive, hassan2009predicting}\end{tabular}    &    \\ \hline \hline
EXP     & Developer experience \cite{mockus2000predicting}  & \multirow{3}{*}{Experience}    \\
REXP    & Recent developer experience \cite{mockus2000predicting}       &    \\
SEXP    & Developer experience on a subsystem \cite{mockus2000predicting} &    \\ \hline
\end{tabular}
\end{table}

\subsection{Semantic Features}\label{semantic}
Semantic features capture the meaning of code text and context with deep learning models. In the last two years, deep learning based techniques, which can automatically extract higher-level features and learn from more complex and high-dimensional data \cite{lecun2015deep}, have been employed for extracting code semantic information. Hoang et al.~\cite{hoang2019deepjit} adopt convolutional neural networks to learn semantic features from code commit messages and code changes for predicting defects. 
Ni et al.~\cite{feng2020codebert} leverage the popular pre-trained model, CodeBERT to extract semantic features of code changes.

 
\subsection{Code Refactoring}
Code refactoring is frequently adopted by developers to make code easier to understand and maintain, improving the code structures without changing their external behavior~\cite{fowler1999refactoring}. Refactoring has been proven useful for relieving code smells \cite{fowler1997refactoring, halepmollasi2022exploring, lacerda2020code}, which notably affects code readability~\cite{sellitto2022toward}. There are studies that investigate the relation between refactoring and code smell \cite{halepmollasi2022exploring, lacerda2020code, fowler1997refactoring}, commit tangling~\cite{herbold2022fine}, and bug-inducing~\cite{di2020relationship, bavota2012does, halepmollasi2022exploring}. 
Bavota et al. \cite{bavota2012does} and Di Penta et al. \cite{di2020relationship} carried out empirical studies and found that commits with code refactoring are more likely to induce bugs, mainly because code refactoring involves changes between classes, methods, and attributes. Also, code refactoring may influence labeling of bug-inducing commits, as revealed by Neto et al., who come up with RA-SZZ for more accurate bug-inducing commits identification~\cite{neto2018impact}. Fan et al. performed an empirical study to learn the impact of refactoring may cause to the labeling of bug-inducing commits \cite{fan2019impact}. Recently, RAT \cite{niu2023rat, niu2024rat} and CodeTracker \cite{jodavi2022accurate} have been proposed to keep track of code entity history based on RefactoringMiner \cite{tsantalis2020refactoringminer, Tsantalis:ICSE:2018:RefactoringMiner}.

However, as far as we know, there is no such work that investigates the impact of code refactoring on SOTA JIT-DP approaches. Furthermore, advanced refactoring mining tools such as RefactoringMiner \cite{tsantalis2020refactoringminer, Tsantalis:ICSE:2018:RefactoringMiner} and RefDiff \cite{silva2017refdiff, silva2020refdiff} are capable of detecting only the code changes directly involved in refactoring, but not the refactoring propagation, which refers to the code changes triggered by the initial refactoring action. Hence, to the best of our knowledge, we are the first to systematically investigate the impact of code refactoring and refactoring propagation on JIT-DP approaches.


\section{Motivating Examples}
\label{sec:motivation}

\subsection{Example of Refactored ``Buggy'' Commit}
This type of example shows a kind of commit that has substantial code changes, predominantly involving code refactoring activities. It has been shown by Nagappan and Ball \cite{nagappan2005use} and Moser et al. \cite{moser2008comparative} that the size of changes, measured by metrics (e.g., the number of added lines of code or the number of modified files), serves as a reliable indicator of bug-inducing modules. Consequently, code change metrics-based approaches might inaccurately classify such commits as \textit{bug-inducing}, when in reality, they are \textit{bug-free}.

\begin{figure}[!htbp]
\centering
\includegraphics[width=\linewidth]{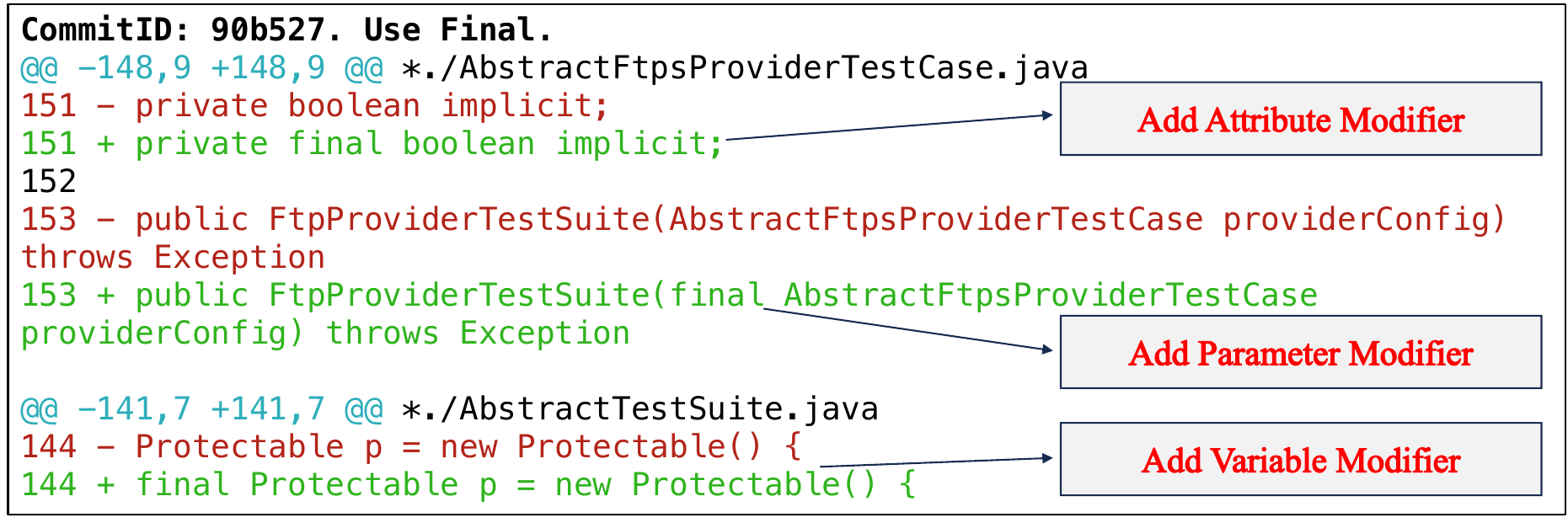}
\caption{Example of Code Refactoring in Commit \href{https://github.com/apache/commons-vfs/commit/90b527}{90b527}.}
\label{fig:motivation}
\end{figure}

There is an example from project commons-vfs,\footnote{https://github.com/apache/commons-vfs/commit/90b527} which has modified 9 files with 20 added lines and 20 removed lines. It appears to be a large commit with a high entropy. JIT-DIL \cite{yan2020just}, a metric-based SDP approach, predicts it as a bug-inducing commit. However, upon closer examination, we notice that the commit consists of 23 code refactoring, encompassing three types, namely, ``Add Parameter Modifier'', ``Add Variable Modifier'', and ``Add Attribute Modifier''. All the reported 20 additions and 20 deletions arise solely from these refactoring. 
Fig.~\ref{fig:motivation} illustrates examples of these three refactoring types in the commit. If executed completely, such refactoring shall not induce bugs.
If the model is aware of code refactoring, it should correctly deduce that the genuine code additions and deletions, exclusive of refactoring, are actually zero, leading to the classification of this commit as clean. Note that the code itself might still be buggy, but that bug was not introduced by the refactoring commit, hence the commit is considered as clean.

The other example of code change from the project commons-math,\footnote{https://github.com/apache/commons-math/commit/7c172a} changed 4 files with 63 additions and 28 deletions. Upon applying refactoring detection on this commit, we have identified all 63 additions are code refactoring of ``Add Method Annotation''. That is, all 63 additions solely involve adding method annotations. Consequently, this kind of code change is unlikely to introduce new bugs. Another example of similar code changes can be accessed from the project commons-validator.\footnote{https://github.com/apache/commons-validator/commit/3ce16c}

\subsection{Example of Refactoring Propagation and Code Tangling}
The function of refactoring detection tools is to detect the existence of refactoring. The fact is that there are still refactoring propagation, as well as changes tangled with refactoring, which are out of the reach of refactoring detection tools. Refactoring propagation does not actually participate in the refactoring process but is merely triggered by it. For instance, if a method is renamed, caller statements also need corresponding modifications. Such code changes should also be considered safe code changes; however, existing refactoring detection tools fail to identify them or do not output them. On the other hand, changes tangled with refactoring may be identified by refactoring detection tools, but they involve other code changes as well. While we can ensure that the refactored part does not introduce bugs, there is still a possibility of bugs being introduced in the edited part. Therefore, for code changes tangled with refactoring, we cannot guarantee that they are safe code changes.

\begin{figure}[htbp]
\centering
\includegraphics[width=\linewidth]{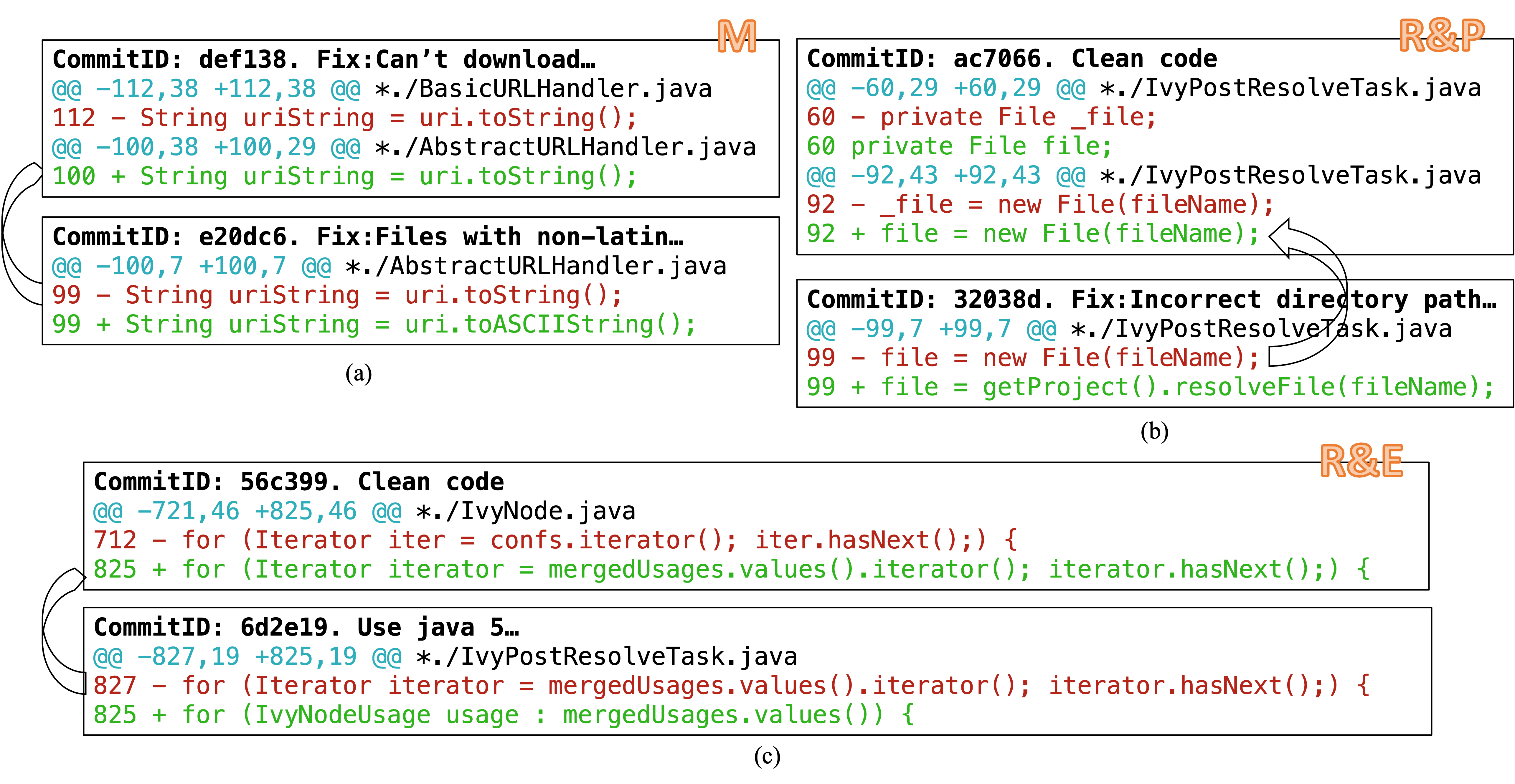}
\caption{Example of Refactoring (a), Refactoring Propagation (b), and Code Tangling (c).}
\label{fig:motivation2}

\end{figure}

In Fig.~\ref{fig:motivation2}, we present examples of moving, refactoring propagation, and code tangling in the ant-ivy project \footnote{https://github.com/apache/ant-ivy}. Example (a) illustrates a bug is fixed in commit e20dc6, while the last revision introducing the buggy line is commit def138. However, in this commit, the buggy line was moved from another file, ``.../BasicURLHandler.java''. We can assert that bugs exist in such refactored code; however, they are never the root cause of the bug. In example (b), ``\_file'' was renamed to ``file'' in line 60, commit ac7066. Consequently, line 92 was updated due to the renaming in the same commit. Later, this line was identified as buggy, and the value was modified in commit 32028d. Nonetheless, refactoring propagation itself is not the root cause of bugs and it is bug-free. In example (c), ``iterator'' was renamed from ``iter'' in commit 56c39, accompanied by a modification in the value. Later, this line was identified to contain a bug in commit 6d2e19, attributed to the modification rather than the refactoring part. Refactoring detection tools can detect refactoring in line 825, but they cannot determine whether line 825 underwent safe refactoring.

In summary, refactoring and refactoring propagated code changes should not be the cause of bugs. When executed correctly, refactoring does not introduce bugs. However, refactoring intertwined with other changes can potentially introduce bugs. This is an area of research that current refactoring detection tools have not yet fully addressed.

\section{CAT: \underline{C}ode Ch\underline{A}nge \underline{T}actics Analysis}\label{sec:approach}

To delve into 1) the impact of code refactoring and refactoring propagation on SOTA JIT-DP approaches, and 2) avenues for enhancing SOTA approaches, we propose CAT, which include 1) fine-grained categories of code changes (Section~\ref{sec:c3})  and 2) strategies for labeling code change categories (Section~\ref{sec:cat}). Leveraging CAT, the data labeling of bug-inducing commits can be augmented (Section~\ref{sec:augment}), and further optimize SOTA JIT-DP approaches (Section~\ref{sec:ra-jitdp}). 

\subsection{Code Changes Categories}\label{sec:c3}
When using \textit{Git diff}, each code change is either labeled as \textit{add} or as \textit{delete} (i.e., $c \in \{add, delete\}$). However, such labels are coarse-grained, because moving a statement would be labeled as \textit{delete} an old line and \textit{add} a new line, even though it is the same statement moving around. Besides, there are also statements being \textit{refactored}, \textit{propagated} by refactoring, or \textit{edited}. Different types of code changes are tangled together within the same commit, and even within the same statement, there can be various code changes mixed together. To untangle such code changes at statement level, we propose fine-grained categories of code changes. The atomic code changes include:

\begin{itemize}
    \item \textit{$<$Add/Del$>$}: A changed line of code only involves adding (deleting) a statement.
    \item \textit{$<$Add/Del$>$\_Move}: A line of code is moved from one place to another place, the content is not modified at all.
    \item \textit{$<$Add/Del$>$\_Refactoring}: A line of code change is part of code refactoring, including rename, extract, and inline.
    \item \textit{$<$Add/Del$>$\_Propagation}: A line of code change is triggered by a code refactoring, for example, renaming the callee (method being called) triggers the modification within the caller (method calling another one).
    \item \textit{$<$Add/Del$>$\_Edit}: Apart from refactoring and propagation, there are other modifications within the line of code.
\end{itemize}

These atomic code change types are intended to differentiate refactoring-related code changes (which have a low risk of introducing bugs) from other complex code changes (which have a high risk of introducing bugs). When implemented correctly, refactoring-related changes do not introduce bugs, while new code additions are more likely to do so. Within the realm of refactoring, ``move'' is a specific type where the content remains unchanged. ``Refactoring'' includes over 90 distinct types of refactoring. ``Propagation'' refers to code changes initiated by refactoring; although these changes are not part of the refactored code, they should also not introduce bugs.

Apart from the above atomic code change types, there can be change tangling within statements, e.g., refactor and edit a statement in the same statement. To this end, there are also composite change types: \textit{$<$Add/Del$>$\_Refactoring\_Propagation}, \textit{$<$Add/Del$>$\_Refactoring\_Edit},\textit{$<$Add/Del$>$\_Propagation\_Edit}, and \textit{$<$Add/Del$>$\_Refactoring\_Propagation\_Edit}. The complete list of categories can be found in Table~\ref{tab:types}.

\begin{table}[htbp]
\caption{Code Change Types of Each Line.}
\label{tab:types}
\begin{tabular}{l|l}
\hline
\textbf{Add}                            & \textbf{Delete}                            \\ \hline
Add                       & Delete                       \\
Add\_Move                 & 
Delete\_Move                 \\
Add\_Refactoring             & 
Delete\_Refactoring             \\
Add\_Propagation            & 
Delete\_Propagation            \\
Add\_Edit                 & Delete\_Edit                 \\
Add\_Refactoring\_Propagation       & Delete\_Refactoring\_Propagation       \\
Add\_Refactoring\_Edit            & Delete\_Refactoring\_Edit            \\
Add\_Propagation\_Edit           & Delete\_Propagation\_Edit           \\
Add\_Refactoring\_Propagation\_Edit & Delete\_Refactoring\_Propagation\_Edit \\ \hline
\end{tabular}
\end{table}

In conclusion, our classification system comprises 18 distinct categories of code change types. These categories are independent and do not overlap with each other. Each statement of code changes within a commit should be labeled as only one corresponding category. It is important to note that our focus is on distinguishing refactoring and refactoring propagation from other code changes to minimize noise in JIT-DP. Therefore, we do not untangle feature requests or bug-related code changes to maintain high accuracy in our classification results. 

\subsection{Code Change Tactics Analysis}\label{sec:cat}
To label categories to each statement in code changes, we propose CAT. CAT takes two revisions of a Java project as input and generates a list of statements that have been updated between these revisions, with each statement categorized into one of the 18 categories. 

The main target of CAT is to identify and separate out refactored and refactoring propagated statements in each commit. For detecting code refactoring, we adopt the SOTA tool RefactoringMiner \cite{tsantalis2020refactoringminer, Tsantalis:ICSE:2018:RefactoringMiner}, which is currently the most comprehensive and accurate refactoring detection tool. However, one limitation of RefactoringMiner is that it only returns statements directly involved in the refactoring process, disregarding other code changes induced by refactoring. For instance, renaming a method may lead to alterations in statements that call this method, which is out of the reach of refactoring detection tools. Moreover, code tangling can also happen at statement level, i.e., code refactoring, refactoring propagation, and statement update can happen in the same statement in one commit, while RefactoringMiner only outputs statements directly involved in code refactoring. The essence is that RefactoringMiner can detect the presence of refactoring, while we require obtaining minimal (i.e., identify only those parts of a change statement that were actually subject to refactoring) and complete refactoring (i.e., include also direct effects of refactoring on other code parts).

\begin{figure}[htbp]
\centering
\includegraphics[width=\linewidth]{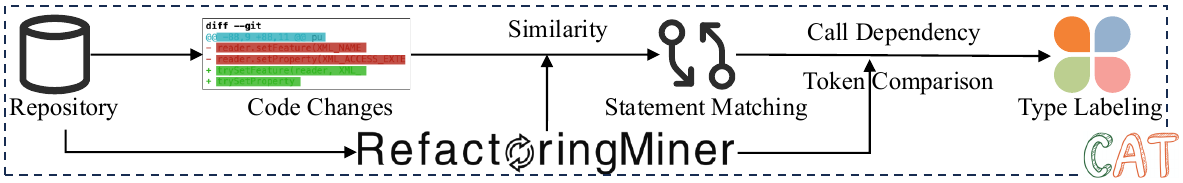}
\caption{Framework of CAT.}
\label{fig:cat}
\end{figure}

The overall framework of CAT is presented in Fig.~\ref{fig:cat}. Firstly, CAT uses \textit{git diff} to obtain the code changes between two adjacent revisions, which is a list of \textit{added} and \textit{deleted} statements. Then, CAT uses RefactoringMiner to identify statements involved in refactoring. If there is refactoring related to class/method/attribute/parameter/variable altering (add, remove, rename), call dependencies will be analyzed to identify refactoring propagation. Statements calling class/method/attribute/parameter/variable name before/after altering will be identified as refactoring propagation. 

The matching of two statements is a core function that we use throughout the labeling process. We employ a two-stage detection process for statement matching. Firstly, our statement matching function utilizes the \textit{JGit.diff.EditList} function to narrow down the scope of matches, which encapsulates the differences or edits (insertions, deletions, replacements) between two versions of a text file. Then, within each \textit{editList}, we look for hints for statement matching from the output of RefactoringMiner. For the remaining statements, we compute the term frequency and cosine similarity of added statements and deleted statements within the same \textit{EditList}. Statements with the highest similarity and higher than 80\% \footnote{In this study, we establish a higher threshold to ensure that the purely refactored lines identified by CAT maintain a high level of precision, allowing us to label them as bug-free in the JIT-DP analysis.} will be matched. For statements that could not be matched, they would be labeled as solely \textit{Add/Del}. We thus ensure that a statement is only labeled as refactored when we are highly certain of it, while some refactoring might go undetected and will be classified as regular changes.

For each pair of match statements, we first utilize the results generated by RefactoringMiner to identify code tokens associated with refactoring. We then compare pairs of statements involving refactoring and refactoring propagation. If there are changes to other code tokens outside of those associated with refactoring, such as additions, removals, and insertions, we also consider this as an edit involved in the modification. Note that \textit{move} is a special type of refactoring, where the two matched statements have exactly the same code tokens, but different positions in the two revisions. It typically happens within \textit{extract}, \textit{inline}, and \textit{move} refactoring.

\subsection{Augmented Data Labeling Leveraging CAT}\label{sec:augment}

Different from RA-SZZ \cite{neto2018impact}, CAT aims at discerning refactoring and its propagated code changes from other types of code modifications. CAT uses the latest version of RefactoringMiner \cite{tsantalis2020refactoringminer}, which can detect more than 90 types of code refactoring. Statements only involve refactoring or refactoring propagation could be identified, i.e., $c \in $ \{\textit{$<$Add/Del$>$\_Move}, \textit{$<$Add/Del$>$\_Refactoring},\textit{$<$Add/Del$>$\_Propagation}, \textit{$<$Add/Del$>$\_Refactoring\_Propagation}\}. 

\begin{figure}[htbp]
\centering
\includegraphics[width=0.8\linewidth]{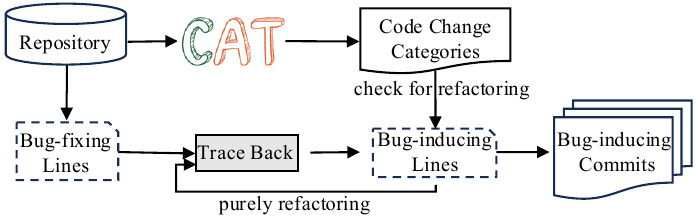}
\caption{Augmented Data Labeling Process.}
\label{fig:datalabel}
\end{figure}

Our augmented data labeling process is shown in Fig.~\ref{fig:datalabel}. Firstly, CAT is applied to detect purely refactored or refactoring propagated statements in commits and saved as $Ref$. Then, in the RA-SZZ algorithm analysis process, CAT starts with identifying bug-fixing commits. We apply $Ref$ to filter out refactoring code changes and retain real bug-fixing statements $bf$. For each statement $s$ $\in$ $bf$, the RA-SZZ algorithm would trace back the last revision $s'$ that induces $s$. If $s'$ is refactored or propagated in that revision (i.e., $s'$ $\in$ $Ref$), then RA-SZZ would continue to trace back the previous revision until that $s'$ $\notin$ $Ref$. Such $S'$ would be bug-inducing code changes $bi$, and the corresponding commit would be labeled as the bug-inducing commit.

Applying CAT to the RA-SZZ algorithm ensures a more accurate bug-inducing dataset for JIT-DP, as CAT is able to distinguish purely refactored and refactoring-propagated code changes from tangled refactoring.

\subsection{JIT-DP Leveraging CAT} \label{sec:ra-jitdp}
\begin{figure}[htbp]
\centering
\includegraphics[width=\linewidth]{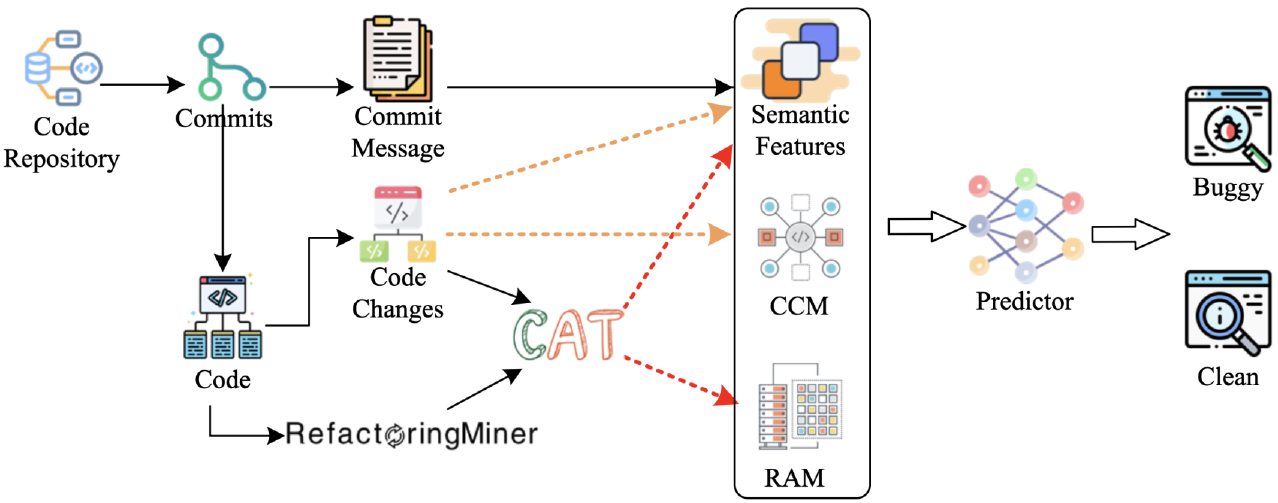}
\caption{Framework of Leveraging CAT in JIT-DP.}
\label{fig:framework}
\end{figure}

In this work, CAT is not only used to enhance bug-inducing commits dataset labeling, but also to improve the performance of JIT-DP approaches. Fig.~\ref{fig:framework} demonstrates the framework for employing CAT into SOTA JIT-DP approaches.

\subsubsection{Code Change Metrics-based JIT-DP Leveraging CAT} \label{sec:cat1}
SOTA JIT-DP approaches mainly adopt the widely used 14 code change metrics proposed by Kaimei et al. \cite{kamei2012large} to extract features of code change. They are defined according to empirical study of code changes. Then machine learning algorithms (e.g., Random Forest) or fully connected layers take the 14-dimensional vector as input to output the defectiveness of each commit. CommitGuru \cite{rosen2015commit} is a widely used online tool designed to automatically extract these 14 code change metrics. However, these 14 metrics are too coarse-grained to capture code changes at the class and method levels, let alone capture refactoring information.

To enhance the state of the art, we come up with more RAMs
(as shown in our replication package), which can capture refactoring-related characteristics of code changes.
The first type focuses on line-level characteristics, with 18 RAMs that calculate the number of code changes for each category in Section~\ref{sec:c3}, i.e., group the code changes according to CAT labeling. The second type addresses class-level features. We calculate the total number of added, removed, moved, refactored, propagated, and updated classes, and the average number of each type of code changes per class. These are an additional 24 features. Similarly, we calculate method-level characteristics, i.e., number of added, removed, moved, refactored, propagated, and updated methods, as well as average number of each type of code change per method, which are 24 more features. 

Overall, there are 66 RAMs proposed. Our RAMs not only capture refactoring-aware features, but also class-level and method-level code change size.
Instead of replacing the original 14 code change metrics, we combine these code change metrics with our new RAMs in the models, since our metrics do not include code history information.

\subsubsection{Semantic Feature-based JIT-DP Leveraging CAT}\label{sec:cat2}
For semantic features, the common practice is separating code changes into ``added lines''  and ``deleted lines'', and then using deep learning techniques (e.g., CodeBERT \cite{feng2020codebert} or Convolutional Neural Network \cite{lecun1989backpropagation}) to generate the corresponding embedding vectors. With CAT, we are able to obtain more detailed categories of code changes as listed in Section~\ref{sec:c3}. 
Thus, instead of distinguishing only between ``added lines'' and ``deleted lines'' we enhance the semantic feature extractors with more code change categories with respect to refactoring. 

Moreover, there is a limit in the length of the input of deep learning models, like CodeBERT-base, which is 512. Sometimes, the code change can be huge and a lot of useful information may be lost due to the length limitation. Now that CAT is able to label code change types, we group all these code changes based on their categories, and then rearrange the inputs of different types of code changes. We place the types of code changes that are more informative (e.g., \textit{Add}, \textit{Add\_Edit}, \textit{Add\_Refactoring\_Propagation\_Edit}) to defect prediction at the front of the input, and less relevant information (e.g., \textit{Add\_Move}, \textit{Add\_Refactoring}, \textit{Del}, \textit{Del\_Move}) at the back. This way, we ensure that useful information is not lost when truncating the input.

\section{Experimental Setup} \label{sec:experimental}

\subsection{Research Questions}
To investigate how refactored lines identified in commits influence JIT-DP, we carry out a comprehensive set of experiments to address the following research questions (RQs):

\begin{itemize}
    \item RQ1: How common are code refactoring and refactoring propagation throughout the history of code evolution?
    \item RQ2: To what extent are SOTA JIT-DP approaches affected by code refactoring and refactoring propagation within code changes?
    \item RQ3: To what extent are we able to improve SOTA JIT-DP approaches leveraging code change labels by CAT?

\end{itemize}

Since the basis of our work is code refactoring and refactoring propagation, the objective of RQ1 is to understand how frequently code refactoring and refactoring propagation occur in the evolution history of code. If little refactoring and refactoring propagation happen throughout the whole history, the impact would be negligible.

Throughout RQ2, we investigate how refactoring information influences the current state of the art in terms of code change metrics and semantic features.

RQ3 aims at seeking the possibility of enhancing SOTA JIT-DP approaches by taking code refactoring into consideration.

\subsection{JIT-DP Models}
We identified six recent SOTA techniques that target JIT-DP with either code change metrics (i.e., Deeper~\cite{yang2015deep}, JIT-DIL~\cite{yan2020just}, LApredict~\cite{zeng2021deep}) or semantic features (i.e., DeepJIT \cite{hoang2019deepjit}), or both (i.e., JITLine~\cite{pornprasit2021jitline}, JIT-Fine~\cite{ni2022best}).

\noindent
(2015) - \textbf{Deeper} \cite{yang2015deep} utilizes 14 code change metrics proposed by Kamei et al. \cite{kamei2012large} to build a deep belief network prediction model for SDP.

\noindent
(2019) - \textbf{DeepJIT} \cite{hoang2019deepjit} builds a defect prediction model with convolutional neural networks to learn semantic features from commit messages and code changes.

\noindent
(2020) - \textbf{JIT-DIL} \cite{yan2020just} leverages 14 code change metrics to build a defect prediction model with logistic regression.

\noindent
(2021) - \textbf{LApredict} \cite{zeng2021deep} predicts defect commits by leveraging ``lines of code added'' with a logistic regression classifier.

\noindent
(2021) - \textbf{JITLine} \cite{pornprasit2021jitline} combines both 14 code change metrics and token features to build a defect prediction model with a random forest classifier.

\noindent
(2022) - \textbf{JIT-Fine} \cite{ni2022best} leverages the combination of 14 code change metrics and semantic features for defect prediction.


We select the above six techniques because they are frequently adopted as baselines or have shown promising results in recent studies.
To avoid unnecessary bias introduced during implementation, we adopt the replication package\footnote{https://github.com/jacknichao/JIT-Fine} provided by Ni et al.~\cite{ni2022best}, which includes five of our baseline approaches. For the last one, JIT-DIL~\cite{yan2020just}, we adopt their open-sourced implementation on GitHub.\footnote{https://github.com/MengYan1989/JIT-DIL} According to the reported result, JIT-Fine performs best among all six baselines so far. For fair comparison, we adopt exactly the same parameters as the original replication package does, but only modify the input of the models.

\subsection{Improving Accuracy of Bug-inducing Commit Dataset}
\begin{table}[]
\caption{Statistics of Studied Dataset.}
\label{tab:dataset}
\resizebox{\linewidth}{!}{%
\begin{tabular}{lcccccc}
\hline
\multicolumn{1}{c}{\multirow{2}{*}{Projects}} & \multicolumn{3}{c}{Commit-level} & \multicolumn{3}{c}{Line-level} \\ \cline{2-7} 
\multicolumn{1}{c}{}                          & All     & Buggy   & Buggy Ratio  & All     & Buggy  & Buggy Ratio \\ \hline
ant-ivy                                       & 1,771    & 350     & 18.7\%       & 14,853   & 1,650   & 11.1\%      \\
commons-bcel                                  & 825     & 64      & 7.3\%        & 1,626    & 310    & 19.1\%      \\
commons-beanutils                             & 611     & 38      & 6.1\%        & 1,724    & 123    & 7.1\%       \\
commons-codec                                 & 761     & 37      & 4.7\%        & 2,074    & 246    & 11.9\%      \\
commons-collections                           & 1,823    & 50      & 2.7\%        & 3,434    & 181    & 5.3\%       \\
commons-compress                              & 1,630    & 179     & 10.9\%       & 7,167    & 627    & 8.7\%       \\
commons-configuration                         & 1,838    & 154     & 8.4\%        & 7,334    & 650    & 8.9\%       \\
commons-dbcp                                  & 1,037    & 54      & 5.6\%        & 2,500    & 192    & 7.7\%       \\
commons-digester                              & 1,079    & 19      & 1.8\%        & 385     & 87     & 22.6\%      \\
commons-io                                    & 1,142    & 71      & 6.4\%        & 2,822    & 196    & 6.9\%       \\
commons-jcs                                   & 831     & 86      & 10.6\%       & 4,796    & 450    & 9.4\%       \\
commons-lang                                  & 2,969    & 155     & 4.9\%        & 6,332    & 563    & 8.9\%       \\
commons-math                                  & 4,026    & 362     & 8.3\%        & 16,960   & 3,046   & 18.0\%      \\
commons-net                                   & 1,121    & 117     & 10.4\%       & 4,834    & 441    & 9.1\%       \\
commons-scxml                                 & 544     & 45      & 8.6\%        & 4,054    & 242    & 6.0\%       \\
commons-validator                             & 598     & 35      & 6.0\%        & 1,189    & 114    & 9.6\%       \\
commons-vfs                                   & 1,110    & 113     & 10.3\%       & 4,902    & 384    & 7.8\%       \\
giraph                                        & 844     & 172     & 19.3\%       & 16,748   & 1,904   & 11.4\%      \\
gora                                          & 553     & 39      & 7.1\%        & 3,397    & 133    & 3.9\%       \\
opennlp                                       & 1,086    & 89      & 8.4\%        & 3,912    & 326    & 8.3\%       \\
parquet-mr                                    & 1,120    & 168     & 14.1\%       & 8,325    & 729    & 8.8\%       \\ \hline
All                                           & 27,319   & 2,397    & 8.6\%        & 119,368  & 12,594  & 10.0\%      \\ \hline
\end{tabular}
}
\end{table}

In a previous study \cite{ni2022best}, the above six baselines have been evaluated on the JIT-Defects4J dataset. This dataset was constructed by two groups of people in two stages. Firstly, Herbold et al. \cite{herbold2022fine} manually created the Line-Labelled Tangled Commits for Java (LLTC4J),\footnote{https://smartshark.github.io/dbreleases/} which only focuses on bug-fixing commits. They hired 45 participants to label each line of code as bug-fixing or other types unrelated to bug-fixing, to accurately locate the bug-fixing lines in commits and to eliminate bias caused by tangled commits. Then, taking LLTC4J as a starting point, Ni et al. \cite{ni2022best} extended the dataset with labeling buggy commits and clean commits, that is, defect-inducing commits. They also extracted the line label in defect-inducing commits. Specifically, for the bug-fixing lines in the LLTC4J dataset, Ni et al. used \textit{git blame} to find the corresponding bug-inducing commits and lines. In this way, the JIT-Defects4J could greatly reduce the scope of defect-introducing candidates and accurately label each line of code. All other commits are treated as clean commits. Next, Ni et al. used PyDriller \cite{spadini2018pydriller} to collect code changes and commit messages of each commit, which are the input of semantic feature. Then, to extract 14 change-level defect features (presented in Table~\ref{tab:14metrics}), Ni et al. adopted CommitGuru \cite{rosen2015commit}, which has been widely used for automatically extracting code change metrics \cite{pornprasit2021jitline, yang2016effort, chen2018multi}. More details of the construction process of JIT-Defects4J are available in the original papers \cite{ni2022best, herbold2022fine}.

Since the bug-fixing commits were labeled manually, we believe that this label has a high level of credibility. However, for the bug-inducing commit labeling, Ni et al. only used \textit{git blame}, which is very likely to label refactored code changes as bug-inducing commits. Thus we decided to start from the second stage to improve the accuracy of bug-inducing commit dataset. In the second stage, when \textit{git blame} returns a statement $s^+$ as bug-inducing, we apply our CAT to check if $CAT(s^+)$ belongs to any of the labels: \textit{Add\_Move}, \textit{Add\_Refactoring}, \textit{Add\_Propagation}, and \textit{Add\_Refactoring\_Propagation} (Only \textit{Add\_*} code changes will be labeled as bug-inducing). If $CAT(s^+)$ $\in$ $\{$\textit{Add\_Move}, \textit{Add\_Refactoring}, \textit{Add\_Propagation}, \textit{Add\_Refactoring\_Propagation}$\}$, we firstly find the corresponding $s^-$ in the same commit. Then we use \textit{git blame} to find the last revision of $s^-$. We repeat the process until the statement is a refactoring-free code change, and it would be labeled as bug-inducing code. The corresponding commit would be labeled as bug-inducing commit.

Finally, we re-annotate 1,676 out of 12,594 bug-inducing lines of code changes, which are actually refactored or refactoring propagated changes. 
We identified a total of 145 clean commits originally marked as buggy commits, and upon further investigation of 1,676 lines of code, we additionally discovered 210 new buggy commits. Among them, 35 commits were not included in the dataset (the JIT-Defects4J dataset excludes commits change over 10,000 lines), so we omitted them. 
The statistics of the relabeled dataset are presented in Table~\ref{tab:dataset}. There are 21 projects with 27,319 commits, among which 2,397 are buggy commits. The buggy ratio of projects varies from 1.8\% to 19.3\%.

To verify the accuracy of CAT, we conduct a manual review of the labeled data. Specifically, we randomly select code changes labeled as $<$\textit{Add\_*}$>$ by CAT, choosing 20 examples from each category for a total of 180 examples. For these cases, we identify the specific code changes and their corresponding deleted lines of code (if applicable). Two of our authors then perform manual labeling of the code change types by examining the code differences. 
Finally, we compare the manually labeled results with those produced by CAT. Our annotations achieved an accuracy of up to 95\%, with the two annotators agreeing on 166 out of 180 samples, resulting in an agreement rate of 92.2\%. The remaining 14 discrepancies were resolved through discussion until a consensus was reached.
Additionally, we manually filtered the re-labeled JIT-Defects4J dataset to ensure the accuracy of our annotations. Specifically, for lines labeled as bug-inducing in the JIT-Defects4J dataset, if CAT marks them as \textit{Add\_Refactoring}, \textit{Add\_Propagation}, or \textit{Add\_Refactoring\_Propagation}, we conduct a manual review by comparing the code differences to determine whether CAT's labeling is accurate. If CAT's labeling is correct, we then trace back to identify the actual bug-inducing line. We find that CAT fails to identify instances where a single line of code is split into two lines. Given the relatively high threshold for statement matching (80\%), CAT may have difficulty identifying matches for code changes that encompass multiple complex refactorings. However, this trade-off is necessary to ensure that lines classified by CAT as purely refactored or propagated retain a high degree of precision. Furthermore, the absence of line numbers in the original buggy line dataset poses challenges in accurately determining the correct code lines within the same commit, particularly when multiple lines of code are identical.

\subsection{Evaluation Measures}
\sloppy
To evaluate the effectiveness of JIT-DP approaches, we adopt commonly used classification evaluation measures (i.e., Precision, Recall, F1-score, and AUC) and effort-aware measures (i.e., Recall@20\%Effort, Effort@20\%Recall, $P_{opt}$).

\subsubsection{Classification Evaluation Measures}
Classification evaluation measures evaluate the performance of classifiers without considering the effort. These measures include: Precision, Recall, F1-score, and AUC. 

\textbf{Precision}: the ratio of correctly predicted commits to all commits predicted as bug-inducing. 

\textbf{Recall}: the ratio of correctly predicted bug-inducing commits to the actual bug-inducing commits. 

\textbf{F1-score}: the harmonic mean of recall and precision ($F1-score = \frac{2\times Recall\times Precision}{Recall+Precision}$)

\textbf{AUC}: the area under the curve of the receiver operating characteristics (ROC) \cite{lessmann2008benchmarking}. AUC ranges from 0 to 1.
\subsubsection{Effort-aware Measures}
Effort-aware measures assess the effectiveness of JIT-DP models in terms of quality assurance effort. Similar to previous work \cite{ni2022best, pornprasit2021jitline, ni2020revisiting}, we choose 20\% of total lines of code as the proxy of inspection effort. These measures include: Recall@20\%Effort, Effort@20\%Recall, and P$_{opt}$.

\textbf{Recall@20\%Effort(R@20\%E)}: measures how many buggy lines can be accurately found when inspecting 20\% of total lines of source code. A high value of this measure indicates that the predictor can rank many buggy lines at the top, and many actual buggy lines can be found given an amount of effort.

\textbf{Effort@20\%Recall(E@20\%R)}: measures how much effort required to find the 20\% actual buggy lines of the whole revision. A low value of this measure means developers spend a small amount of effort to find the top 20\% actual buggy lines.

\textbf{P$_{opt}$}: is a threshold-free evaluation measure. It is calculated as $1-\frac{Area(Optimal)-Area(M)}{Area(Optimal)-Area(Worst)}$. In the optimal and worst models, commits are sorted in decreasing and ascending order by defect densities, respectively. Area(M) represents the area under the curve corresponding to the model M.

\section{Results and Discussion}
\label{sec:results}

\subsection{RQ1: Frequency of Code Refactoring and Refactoring Propagation in Code History} \label{sec:rq1}

To investigate the frequency of refactoring and refactoring propagated code lines appearing throughout the software lifecycle, we utilize CAT to categorize the types of code changes in each commit of the dataset, and we then tally the distribution of different categories of added lines in each project, since removed lines are thought to be bug-free \cite{sliwerski2005changes}. The results are presented in Table~\ref{tab:category}. 327 commits with over 1000 lines of code change were excluded from this table to avoid bias from large changes.

The results show that there are 1,102,757 lines purely added (Column ``Add''), 156,794 lines purely edited (Column ``Edit''), 122,774 lines purely refactored (Column ``Refactored'', includes purely move and purely refactored), 10,245 lines propagated (Column ``Propagated'', includes purely propagated and tangle of refactoring and propagation), and 19,454 lines tangled edit with refactoring (Column ``Tangled'', includes move, refactoring and propagation). The vast majority of code changes are simply line additions (78.1\%), followed by pure code edits (11.1\%). Lines purely refactored and moved account for 8.7\% of all the changes, propagated lines account for 0.7\%, and 1.4\% involve a mixture of refactoring and editing. In total, 10.8\% of the added lines are involved in refactoring. CAT enables finer-grained identification of purely refactored code changes. However, relying solely on the results of refactoring detection tools, researchers may easily overlook approximately 0.7\% of propagated code changes and conflate the 1.4\% of tangled changes with purely refactored changes, resulting in a 15.8\% false negative rate in finding refactored code. This is not a trivial figure and can introduce considerable bias to evaluations.

The last column (Column ``Refactored Commits'' ) shows the number of commits that involve at least one refactoring in each project. The results show that among all 21 projects, the amount of refactored commits is between 228 and 1789, which covers 31.2\% to 54.4\% of all the commits. On average, 41.3\% of the commits in our studied dataset involve refactored code.

\begin{table}[]
\caption{Distribution of Code Change Categories across Commits.}
\label{tab:category}
\resizebox{\linewidth}{!}{%
\begin{tabular}{lllllll}
\hline
Project               & Add     & Edit   & Refactored & Propagated & Tangled & \begin{tabular}[c]{@{}l@{}}Refactored\\ Commits\end{tabular} \\ \hline
ant-ivy               & 58,089   & 7,817   & 8,515             & 1,686     & 1,763             & 731            \\
commons-math          & 194,778  & 27,649  & 27,501            & 1,755     & 4,297             & 274            \\
opennlp               & 42,653   & 5,249   & 3,588             & 175      & 567              & 228            \\
parquet-mr            & 57,638   & 5,576   & 8,151             & 387      & 1,171             & 310            \\
commons-lang          & 100,399  & 16,232  & 7,971             & 1,421     & 1,344             & 779            \\
commons-net           & 31,823   & 6,210   & 2,228             & 192      & 265              & 655            \\
commons-collections   & 88,299   & 15,005  & 10,138            & 1,049     & 1,624             & 887            \\
commons-beanutils     & 35,533   & 2,794   & 2,062             & 51       & 296              & 423            \\
commons-codec         & 23,729   & 6,268   & 1,904             & 281      & 630              & 337            \\
commons-compress      & 45,243   & 6,165   & 5,443             & 441      & 957              & 438            \\
commons-configuration & 86,484   & 6,829   & 8,064             & 280      & 1,135             & 419            \\
commons-digester      & 38,860   & 3,741   & 1,621             & 148      & 444              & 964            \\
commons-jcs           & 44,032   & 7,874   & 5,734             & 332      & 833              & 1,789           \\
commons-io            & 41,160   & 6,514   & 3,701             & 234      & 471              & 338            \\
commons-scxml         & 21,914   & 2,310   & 2,712             & 296      & 454              & 256            \\
commons-validator     & 21,254   & 4,647   & 1,395             & 114      & 193              & 191            \\
commons-vfs           & 32,379   & 5,224   & 4,617             & 405      & 754              & 514            \\
giraph                & 68,903   & 5,077   & 10,208            & 341      & 984              & 459            \\
commons-bcel          & 14,639   & 5,695   & 2,093             & 225      & 461              & 233            \\
commons-dbcp          & 24,482   & 4,837   & 2,517             & 380      & 461              & 467            \\
gora                  & 30,466   & 5,081   & 2,611             & 52       & 350              & 596            \\ \hline
Total                 & 1,102,757 & 156,794 & 122,774           & 10,245    & 19,454            & 11,288 \\ \hline
\end{tabular}
}
\end{table}

\begin{center}
\fcolorbox{black}{gray!10}{\parbox{1.0\linewidth}{Refactoring has an extremely common occurrence. Around 41.3\% commits include code refactoring, and 10.8\% of the added lines are part of refactoring changes. This opens up opportunities for refactoring related research.}}
\end{center}

\subsection{RQ2: Impact of Refactoring and Refactoring Propagation on SOTA Approaches} \label{sec:rq2}

\textbf{Experiment Design}: 
The six SOTA approaches were previously evaluated on JIT-Defects4J dataset \cite{ni2022best}, while the dataset ignored refactoring and its propagation during labeling bug-inducing commits. In this work, we augment the JIT-Defects4J dataset with CAT by correcting the label of 320 commits.
In order to investigate the impact of mislabeling, we evaluate the six approaches with the augmented dataset. Specifically, we train Deeper \cite{yang2015deep}, JIT-DIL\cite{yan2020just}, LApredict\cite{zeng2021deep} with code change metrics, and train DeepJIT \cite{hoang2019deepjit} with semantic feature. For JITLine \cite{pornprasit2021jitline}, and JIT-Fine \cite{ni2022best}, we run the code change metrics-based part and the semantic feature-based part separately to obtain more intuitive experimental results. Following Ni et al. \cite{ni2022best}, we choose the same 80\% commits as the training set and the other 20\% commits as the test set. We leverage their code change metrics generated by CommitGuRu \cite{rosen2015commit}.

\textbf{Results}:
The experimental results are presented in Table~\ref{tab:rq2}. The term ``Orig.'' refers to the original JIT-Defects4J dataset, whereas ``Aug.'' denotes our augmented dataset with CAT. Apart from JITLine and JIT-Fine approach, no significant differences are observed in the experimental results. This suggests that existing approaches still demonstrate relatively robust fault tolerance. A mislabeling rate of around 10\% at the commit level does not cause substantial disruption to the model. However, for the JIT-Fine approach based on code change metrics, correcting the dataset results in higher precision in predicting buggy commits but a notable decrease (by 69.7\%) in recall. The F1-score and AUC are increased by 5.2\% and 3.3\%, respectively. Conversely, the semantic feature-based JIT-Fine approach shows significant improvements, with precision, recall, F1-score, and AUC increasing by 27.3\%, 42.8\%, 37.3\%, and 5.3\%, respectively. That of JITLine approach also improved by 25.4\%, 11.4\%, 18.6\%, and 1\%, respectively. More notably, the effort-aware metrics of the semantic feature-based JIT-Fine also show significant improvement. Recall@20\%Effort increased by 74.4\%, Effort@20\%Recall decreased by 75.6\%, and P$_{opt}$ improved by 11.9\%. This is because semantic feature-based models use code changes as inputs, and they can effectively identify refactored code lines and accurately label buggy lines. However, using mislabeled datasets can mislead semantic feature-based models, thereby weakening the models' performance. This further highlights that utilizing accurately labeled datasets can lead to a more balanced evaluation of the models' performance, especially for semantic feature-based models.

Among all six approaches, the precision values are concentrated between 13.8\% and 34\%, with most F1-score clustering around the 20\% mark, while recall values are dispersed between 11.1\% and 71.8\%. DeepJIT achieves a recall as high as 71.8\%, but its precision is only 13.8\%. The reason could be due to the approach only selects 10 lines of code, with code line lengths padded to 512. This would lead to the loss of code information from large code changes and introduce a lot of noisy data. The LApredict approach only utilizes the number of added lines of code as input. Although it performs well in its original dataset, it does not work on the JIT-Defects4J dataset. The JITLine semantic-based approach achieves the highest F1-score at 29.2\%, while the JIT-Fine semantic-based approach attains the highest AUC at 78\% on the corrected dataset. This further demonstrates the exceptional performance of semantic features; it is insufficient to rely solely on traditional code change metrics; the content of the code changes must also be considered.

\begin{center}
\fcolorbox{black}{gray!10}{\parbox{1.0\linewidth}{Code change metrics-based baselines show strong fault tolerance to refactoring-induced mislabeling, with no notable performance drop. In contrast, semantic-based approaches like JITLine and JIT-Fine significantly improve after dataset correction, outperforming all other models. This highlights their ability to distinguish refactored code and more accurately identify buggy lines.}}
\end{center}

\begin{table}[htbp]
\caption{Impact of Refactoring on JIT-DP Approaches (\%).}
\label{tab:rq2}
\resizebox{\linewidth}{!}{%
\begin{tabular}{lcccccccc} 
\hline
Models                      & Dataset      & Precision & Recall & F1 & AUC    & R@20\%E & E@20\%R & P$_{opt}$   \\ \hline
\multirow{2}{*}{LApredict} & Orig.      & 45.5     & 3.2  & 5.9    & 69.4  & 62.5   & 2    & 81.4  \\
                           & Aug.     & 41.7     & 3.3  & 6.2    & 69   & 62.8  & 1.9   & 81.6  \\ \hline
\multirow{2}{*}{Deeper}    & Orig.      & 17.5     & 42.3  & 24.6    & 68.2  & 63.8   & 2.1   & 82.7  \\
                           & Aug.     & 16.6     & 41.2  & 23.7    & 68   & 63.9   & 2    & 82.8  \\ \hline
\multirow{2}{*}{JIT-DIL}   & Orig.       & 17.6     & 59.6  & 27.2    & 72.2  & 64.6  & 1.7  & 82.7  \\
                           & Aug.      & 16.8     & 58.6  & 26.1    & 71.8 & 64.4   & 1.7   & 82.8  \\ \hline
\multirow{2}{*}{JITLine-M} & Orig.       & 23.5     & 10.7  & 14.7    & 67.2  & 61.7   & 2.3   & 81.8  \\
                           & Aug.      & 25.4     & 11.1  & 15.4    & 67.7  & 62.6   & 2.7   & 81.7  \\   \hline
\multirow{2}{*}{JIT-Fine-M} &  Orig.      & 14.9     & 49.9  & 23     & 66.1  & 63.2   & 2    & 82.5  \\
                           &  Aug.     & 20.6    & 29.4 & 24.2   & 68.3 & 63.9  & 1.9  & 82.6 \\ \hline
\multirow{2}{*}{DeepJIT}   & Orig.       & 13.7   & 71.4 & 23  & 71.3 & 62.1 & 1.62 & 83.6 \\
                           & Aug.      & 13.8   & 71.8 & 23.1 & 71.2  & 62.5 & 1.76 & 83.5 \\ \hline
\multirow{2}{*}{JITLine-S} & Orig.      & 22.6   & 27  & 24.6 & 74   & 66.9  & 1.4  & 85.2 \\
                           & Aug.      & 28.3    & 30.1  & 29.2  & 74.8 &  67.3 & 1.4 & 85.4      \\ \hline
\multirow{2}{*}{JIT-Fine-S} & Orig.       & 26.7 & 15.2 & 19.3 & 74.1  & 36.4  & 9   & 76.3 \\
                           & Aug.      & 34   & 21.7 & 26.5 & 78   & 63.5  & 2.2  & 85.4 \\ \hline   
\end{tabular}
}
\end{table}

\begin{table}[htbp]
\caption{Performance of Leveraging CAT to JIT-DP Approaches (\%). }
\label{tab:rq3}
\resizebox{\linewidth}{!}{%
\begin{tabular}{lccccccc} 
\hline
Models     & Precision & Recall & F1 & AUC    & R@20\%E & E@20\%R & P$_{opt}$   \\ \hline
LApredict  & 50$\uparrow$ & 2.9  & 5.4    & 68.1  & 62.4   & 1.9$\uparrow$   & 81.7$\uparrow$  \\
Deeper     & 16.1     & 50$\uparrow$    & 24.4$\uparrow$    & 69.8$\uparrow$  & 63.7   & 2    & 83$\uparrow$ \\
JIT-DIL    & 18.8$\uparrow$     & 56.6  & 28.3$\uparrow$    & 73.6$\uparrow$  & 64.4   & 1.6$\uparrow$   & 83.9$\uparrow$  \\
JITLine-M  & 28.4$\uparrow$     & 15.9$\uparrow$  & 20.4$\uparrow$    & 69.9$\uparrow$  & 60.4   & 2.1$\uparrow$   & 82.1$\uparrow$  \\
JIT-Fine-M  & 23.1$\uparrow$    & 36.4$\uparrow$ & 28.3$\uparrow$  & 70.2$\uparrow$  & 64$\uparrow$    & 1.9  & 82.9$\uparrow$ \\ \hline
DeepJIT    & 13.8 & 76.2$\uparrow$ & 23.4$\uparrow$ & 71.5$\uparrow$  & 64$\uparrow$   & 1.44$\uparrow$ & 83.6$\uparrow$\\
JITLine-S  & 28  & 31.3$\uparrow$ & 29.6$\uparrow$  & 75$\uparrow$    &  67.3     &  1.4         & 85.4      \\
JIT-Fine-S  & 30.2 & 27.7$\uparrow$ & 28.9$\uparrow$ & 78.5$\uparrow$  & 65.7$\uparrow$  & 2.2  & 85.8$\uparrow$ \\ \hline
\end{tabular}
}
\end{table}

\subsection{RQ3: Performance of Applying CAT in JIT-DP} \label{sec:rq3}

\textbf{Experiment Design}: 
To enhance code change metrics-based approaches with CAT, as described in Section~\ref{sec:cat1}, we combine the 66 RAMs
with the original 14 metrics (Table~\ref{tab:14metrics}). There are 80 code change metrics in total. We apply these code change metrics to Deeper \cite{yang2015deep}, JIT-DIL\cite{yan2020just}, LApredict\cite{zeng2021deep}, JITLine \cite{pornprasit2021jitline} and JIT-Fine \cite{ni2022best}. 

For semantic feature-based models, we group the code changes in each commit by code change categories labeled with CAT. Then we adjust the input order of different groups, placing additions, edits, and tangled code changes at the front of the input, while pure refactoring and propagated code changes are placed at the end of the input, to ensure that more useful information is retained when the code is truncated due to length limit.

Moreover, we also employ a very simple filter. Since purely refactored or propagated code changes are bug-free if implemented correctly, so if a commit contains only purely refactored or purely propagated code changes, then that commit should be a clean commit. Therefore, before the model makes predictions, we pre-label commits of this type to further improve the model's effectiveness.

We evaluate these enhanced models on the same training set and evaluation set with RQ2 on the augmented dataset.

\textbf{Results}:
The results of our CAT-enhanced baselines are illustrated in Table~\ref{tab:rq3}. Compared with the results on the augmented dataset in Table~\ref{tab:rq2},
the experimental results suggest that across all baselines, there is an enhancement in accuracy, observed in both code change metrics-based and semantic feature-based approaches. except for LApredict. LApredict demonstrates poor performance, because it relies solely on the ``lines of code added'' feature. Even after filtering out added lines related to refactoring, the model's effectiveness remains insufficient. For the other approaches, F1-score increases range from 1.3\% for DeepJIT to 32.5\% for JITLine-M. All models exhibit enhancements in AUC, with increases between 0.27\% and 3.3\%. Following the incorporation of refactoring information, most models show notable improvements in recall; JITLine-M leads with a 43.2\% increase, while JIT-Fine-M and JIT-Fine-S improve by 23.8\% and 27.7\%, respectively. For JIT-DP, improving recall means that the models can identify more buggy code at an early stage, which is crucial for effective bug detection. In terms of effort-aware measures, Recall@20\%Effort presents mixed results, with some models improving (JIT-Fine-M, DeepJIT, and JIT-Fine-S) and others declining (LApredict, Deeper, and JITLine-M). Conversely, Effort@20\%Recall shows consistent improvements for JIT-DIL (5.9\%), JITLine-M (22.2\%), and DeepJIT (18.2\%), while remaining unchanged for Deeper, JIT-Fine-M, JITLine-S, and JIT-Fine-S. All models also show varying degrees of improvement in P$_{opt}$, up to 1.33\% for JIT-DIL. Particularly noteworthy advancements are observed in code change metrics-based JITLine and JIT-Fine, showcasing increases in their F1-scores by 32.5\% and 16.9\%, respectively. Overall, the improvements in code change metrics-based models are more pronounced than those in semantic feature-based models, leading to significant enhancements in the precision and recall of most models. This indicates that our RAMs indeed capture useful information and are generally effective.

Among all baselines, JITLine and JIT-Fine achieve the best experimental results. Furthermore, after incorporating refactoring information, we further improve JIT-Fine by 9.1\% and 16.9\%, in F1-score, respectively. The recall values increased by 23.8\% and 27.6\%, respectively. This implies that we can identify more buggy commits.
Although our improvements are far from reaching practical applicability, we have still made significant advancements over all SOTA approaches. This suggests that refactoring information cannot be overlooked. In future research, researchers should not overlook this crucial information; rather, they should endeavor to explore improved approaches for utilizing refactoring information.

\begin{center}
\fcolorbox{black}{gray!10}{\parbox{1.0\linewidth}{The integration of CAT improves accuracy in six state-of-the-art models, particularly in recall, underscoring the value of utilizing refactoring information. Moreover, our approach enhances JIT-Fine accuracy by 16.9\% and 9.1\% in F1-score, highlighting its effectiveness in optimizing performance.}}
\end{center}


\section{Threats to Validity} 
\label{sec:threats}

In this section, we discuss the internal and external threats to validity for our work and how we mitigated them. 

\textbf{Internal Validity:} 
The first threat to internal validity is the implementation of the JIT-DP Models. To mitigate this threat, we used the implementations provided by Ni et al.~\cite{ni2022best} and Yan et al.~\cite{yan2020just}, which are also used in related work. 
To reduce the threat brought by dataset, we use JIT-Defects4J \cite{ni2022best}, whose bug-fixing commits are labeled manually.
CAT strongly depends on refactoring information; we applied the RefactoringMiner tool, which is widely used in the literature and has a high recall (94\%) and precision (99.6\%)~\cite{tsantalis2020refactoringminer, Tsantalis:ICSE:2018:RefactoringMiner}. 
Lastly, our results and analysis are dependent on the quality of dataset, which can be a potential threat. We choose the JIT-Defects4J dataset created by Ni et al. \cite{ni2022best}, which leverages manually labeled bug fixing commits. Another internal validity of our research could be that CAT model may not be able to identify all pure refactoring code changes, such as some line-breaking code changes. However, we ensure the reliability of our identified results through code token comparison. 

\textbf{External Validity:} 
One threat to the external validity is that we evaluated only Java projects, although Java is among the most widely used programming languages. The core idea of CAT is to distinguish between different types of code changes, with a particular focus on disentangling code refactoring and its propagation from other modifications. With appropriate refactoring mining tools, CAT can be readily generalized to other programming languages, such as Python and C++.
Another potential threat is that our investigation focuses on open-source projects. The nature of refactoring and bug prediction in closed-source projects is not necessarily the same as in open-source ones. However, we cover a diverse set of 21 open-source projects with differing sizes/domains, and all key findings were uniform.

\section{Conclusions}
\label{sec:conclusion}

In this study, we introduce CAT, the first tool designed to distinguish purely refactoring changes and their propagation from other code changes, with the latter being more likely to introduce bugs. Using CAT, we 1) enhance the accuracy of the JIT-Defects4J dataset for JIT-DP; 2) empirically assess the effects of ignoring refactoring on six SOTA approaches; and 3) improve the performance of JIT-DP approaches by utilizing CAT. Our findings indicate that refactoring is a widespread phenomenon, and misclassifying refactoring as buggy reduces the performance of certain models, particularly semantic feature-based ones. Additionally, incorporating refactoring analysis can significantly improve the performance of SOTA approaches. 

In the future, we will seek more effective methods for utilizing refactoring information. We will also explore the application of large language models for defect prediction. Furthermore, our CAT tool can be applied not only to defect prediction but also to enhance other defect-related research areas, such as bug localization and program repair. We look forward to exploring more possibilities ahead.



\bibliographystyle{IEEEtran}

\input{output.bbl}

\end{document}